%% file: Tikhonov.tex
\documentclass[
 aps, pra,
 amsmath,amssymb,
 11pt,
 final,
tightenlines,
 twoside,
 twocolumn,
 nofloats,
nofootinbib,
 superscriptaddress,
showkeys,
showkeywords
 ]
{revtex4}
\usepackage[T2A]{fontenc}
\usepackage[utf8x]{inputenc}
\usepackage[russian,english]{babel}
\usepackage{graphicx} 
\usepackage{dcolumn} 
\usepackage{bm} 
\usepackage{xcolor}
\input{maik.rty}
\setcitestyle{authoryear,round}
\input{sao_cmd_author.tex}

\begin{document}
\selectlanguage{english}
\keywords{stellar photometry of galaxies: distances to galaxies: brightest stars in galaxies: brightest star method }

\title{BRIGHTEST STARS OF IRREGULAR AND LOW-MASSIVE SPIRAL GALAXIES\footnote{Based on observations with the NASA/ESA Hubble Space Telescope, obtained at the
 Space Telescope Science Institute,
 which is operated by AURA, Inc. under contract No. NAS5-26555. These 
 observations are associated with the proposals 5091, 5375, 5397, 5427, 5915, 5971, 5972,
6431, 6549, 6584, 6695, 6865, 7202, 7496, 8059, 8122, 8192, 8199, 8584,
8601, 9042, 9086, 9162, 9765, 9771, 9774, 9820, 10182, 10210, 10235, 10402,
10433, 10438, 10505, 10523, 10585, 10605, 10696, 10765,  10877, 10885,
10889, 10905, 10915, 10918, 11229,  11307, 11360, 11575, 11718, 11986, 12196,
12546, 12878, 12880, 12902, 12968, 13357, 13364, 13442, 13750, 14678, 15133,
15243, 15275, 15564, 16075.}}

\author{\firstname{N.~A.}~\surname{Tikhonov}}
 \affiliation{\saoname}

 \author{\firstname{O.~A.}~\surname{Galazutdinova}}
 \affiliation{\saoname}
 
  \author{\firstname{G.~M.}~\surname{Karataeva}}
 \affiliation{St.~Petersburg University, St.~Petersburg, 199034 Russia}
 
 \author{\firstname{O.~N.}~\surname{Sholukhova}}
 \affiliation{\saoname}

 \author{\firstname{V.~D.}~\surname{Ivanov}}
 \affiliation{European Southern Observatory, D-85748 Garching bei Munchen, Germany}
 
 \author{\firstname{A.~}~\surname{Valcheva}}
 \affiliation{University of Sofia ''St. Kliment Ohridski'', BG-1164, Sofia, Bulgaria}
 
 \author{\firstname{P.~L.}~\surname{Nedialkov}}
 \affiliation{University of Sofia ''St. Kliment Ohridski'', BG-1164, Sofia, Bulgaria}

\begin{abstract}

A search for a correlation between the luminosities of the brightest stars
and luminosities of their host galaxies was carried out on archived Hubble
Space Telescope (HST) $F606W$ or $F555W$ ($V$) and $F814W$ ($I$) images of about
150 nearby galaxies. The sample contains only galaxies with on-going star
formation (SF) and with known distances we derived with the TRGB-method.
We correlated the average absolute luminosities of the three brightest blue
and the three brightest red stars with the luminosity of a host. We found a
linear relation for both the blue and the red stars in irregular and low-mass
spiral galaxies. Their scatters are sufficiently small ($0\,.\!\!^{\rm m}4$) to make this
relations useful for distance determination for low-mass galaxies.
We found that all 31 dwarf galaxies ($M_B$$>$$-$13$^{\rm m}$) in our sample
lack bright stars ($M_V{\rm(BS)}< -7\,.\!\!^{\rm m}0$), probably due to the physical
conditions that prevent their birth. For galaxies with higher
luminosity in the range $-$18$^{\rm m}$$<$$M_B$$<$$-$$13$$^{\rm m}$, there is
an asymmetry in the distribution of the number of galaxies relative to the
linear dependence, indicating an increase in the fraction of galaxies with
bright stars.
\end{abstract}

\maketitle

\section{INTRODUCTION}

In spiral and irregular galaxies there are continuous processes of star
 formation, sometimes very active. Violent star formation is most often
 observed in galaxies when they interact with each other at close distances.
 The mass distribution of young stars is determined by the Salpeter
 function~\citep{Sal1955}, but in this work we are only interested in the
 most massive and brightest of them. At present, it is possible, with the
 data provided by the Hubble Space Telescope (HST), to study the stellar
 content of many galaxies, to discover the brightest stars in them and
 to find a correlation between the parameters of such stars and host galaxies.
 This question is not simple, since galaxies differ greatly in morphology,
 luminosities and masses, and in addition, they have different metallicities
 of the gas from which stars are born, or the galaxies can be part of various
 spatial structures, groups or clusters, which affects star formation
 processes. In this paper, we consider only the correlation between the
 luminosities of the brightest stars and their host galaxies.

Interest in the brightest stars is due to the fact that in modern astrophysics
 the question of the upper limit on a star's mass has not been resolved yet.
 Massive stars evolve very quickly, and the final stage of a star's evolution
 can be a supernova explosion, the formation of a black hole or neutron star,
 or even the merger of such relativistic objects with an enormous energy
 release. What will be the path of evolution of a star depends on its initial
 mass. Stellar evolution models suggest that initial stellar masses can be as
 high as $500M_{\odot}$~\citep{Yus2013} and more, but so far, stars of much
 smaller mass have been discovered~\citep{Cro2010, Teh2019, Bes2020}.

The theory predicts that the brightest and most massive stars should have a
 low metallicity, that is, they should be born in dwarf galaxies with low
 metallicity and violent star formation. As a rule, such galaxies are
 interacting. However, the observations indicate that most of the real
bright massive stars are located in spiral galaxies (Milky Way, M\,31, M\,33),
and only a few in irregular galaxies like NGC\,6822, SMC, DDO\,68,
IC\,10 \citep{Wof2020}. Addressing the discrepancy between the theory and
observations requires a systematic search and study of bright massive stars
in galaxies with various physical parameters.

Interest in the correlation between the luminosities of stars and host
 galaxies is also in testing the possibility of using the brightest stars to
 determine the distances to galaxies. This method was proposed by Lundmark
 in 1919 \citep{Lun1919}, but only in 1936 did Hubble determine the average
 luminosities of the brightest stars for 145 nearby galaxies \citep {Hub1936}.
 The method was actively used in the 1960s--1990s \citep{Hol1950,San1974,
Vau1978,Hum1983,Kar1994}.

 At present, accurate results on measuring the
 distances to galaxies located no further than 25~Mpc are obtained using the
 TRGB method (Tip of Red Giant Branch) \citep{Lee1993} based on data from the
 Hubble Space Telescope. For more distant galaxies, several methods are used.
 The Tully--Fisher method \citep{TF1977} is widely used for spiral galaxies
 with an  accuracy of approximately $\pm0\fm4$ \citep {Willick1996}. The SNIa
 supernova method has a low accuracy, but \citet{Riess1996} it has been
 increased from $\pm0\fm65$ to $\pm0\fm12$ by introducing additional
 corrections. The work~\citet{Antipova2020} indicates the accuracy of this
 method in $\pm0\fm18$. A big problem
 with this method is that supernovae are rare and have been observed in a
 limited number of galaxies. The Cepheid method is an accurate method, but
 this method has been applied to measure a small number of galaxies, since
 it requires a series of images separated in time and obtained in several
 filters. It can be stated that for distant galaxies today there is no simple
 and accurate method for measuring distances similar to the TRGB method.

The launch of the HST space telescope, the creation of software for automatic
 photometry of stars: DAOPHOT~\citep{Ste1987} and DOLPHOT~\citep {Dol2016},
 and the advent of the TRGB method for determining distances to galaxies
 pushed the method of the brightest stars into the background.

Historically, the brightest stars in a galaxy were used as distance indicators
before the TRGB method, but this early development meant that data came from
photographic plates and with relatively poor angular resolution. Here we
investigate if this technique can be approved. Our effort is driven by a
number of considerations. First, the superb angular resolution of the HST and
the development of software for automatic photometry of large data sets with
severely crowded stars~\citep[DAOPHOT, DOLPHOT; ][]{Ste1987,Dol2016} allow 
expanding the samples beyond what was possible before.
Second, we attempt to introduce additional parameters into the relation
between the luminosity of galaxies and their brightest stars, to account for
effects like metal abundance and age. We investigate if this reduced the
intrinsic scatter of this relation. The improvements can make the brightest
stars method relevant for studies of spiral and irregular galaxies beyond
20--25\,Mpc, where red giants are not visible in HST images, as well as for
very low-mass dwarf galaxies, in which only a small number of blue stars
are observed.

\section{ SEARCH FOR THE BRIGHTEST STARS IN THE GALAXIES }

To compare the luminosities of stars and host galaxies, we need to find the
 brightest stars in the galaxies. At first glance, the task seems easy, since
 bright massive stars (hypergiants) stand out well against the background of
 other objects. However, most dwarf galaxies simply do not have such stars.
 The lifetime of very massive objects, which during this period are the
 brightest stars in galaxies, is extremely short (1--3 million years), and
 the probability of their appearance in low-mass galaxies is small. According
 to the law \citet{Sal1955} of stellar masses distribution, at the birth of
 one star with a mass of $100M_{\odot}$, several thousand stars of lower
 masses should appear. Even in the spiral galaxies (M\,31, M\,33, NGC\,2403)
 with numerous star-formation regions, less than 10 bright stars with masses
 of 150--300 $M_{\odot}$ are known~\citep {Wof2020, Ric2018}.

The second reason, which makes it difficult to distinguish the brightest
 stars of other galaxies, is the superposition of the background stars of
 our Galaxy on the studied fields. The vast majority of background stars have
 a color index \mbox{$(V-I) = 0\fm8$--$1\fm2$}, but sometimes blue background
 stars do not differ in color from young stars of the studied galaxy, which
 complicates the search for really bright stars. An example is the
 metal-poor dwarf galaxy DDO\,68, where, in addition to the bright massive
 LBV star (Luminocity Blue Variable), a brighter blue background
 star~\citep{Tik2021} is visible. That is, the probability for background
 objects to fall into the sample of candidates for bright massive stars is
far from zero.

The third reason, which makes it difficult to search~--- insufficient angular
 resolution of the telescope. The most massive stars are born in clusters
 \citep{deW2004} and are more likely~---but not exclusively~--- to be
 found still residing there, so searches for the brightest stars should
 concentrate on young star clusters, where the crowding and the
 contamination problems are more severe than in the field. In images
 obtained with ground-based telescopes, it is not always possible to
 resolve groups of massive stars into separate stars, even in our Galaxy.
 The use of the Hubble Space Telescope dramatically improves the angular
 resolution of images, but it is not enough for distant galaxies.

The difficulty of distinguishing bright stars from other objects of the galaxy
 is also evidenced by the fact that \citet{San1958} in the images of higher
 quality than those previously available from E.~Hubble showed that E.~Hubble took
 star clusters and H\,II regions for individual stars in the M\,101 galaxy.
 To identify bright hypergiants one can use the property of their brightness
 variability due to the instability of stars of this type, as well as the
 presence of a strong H$\alpha$ emission line in the spectra. However, such
 a detailed research of the stellar properties will require a long study of
 each galaxy, so the task of finding the brightest stars in several hundred
 galaxies will prove impossible.

Taking into account the difficulties listed above, we proceeded from the
 observational facts that most of the bright massive stars found so far stand out
 well from the rest of the stars in the neighborhood. This means that if
 such a star is present in the image, then the stellar photometry package will
 determine its brightness and color, and using these parameters, the star
 will be found on the CM diagram of the stars.
 The stars selected for their high luminosity and low color index were
 reviewed in images with different levels of visualization. This made it
 possible to exclude very compact star clusters, which had a nearly stellar
 photometric profile of the image. Compact H\,II regions were detected by
 their low color index and different photometric profiles in the $F814W$ and
 $F606W$ bands. The stars selected in this way have compiled our list of
 the brightest blue stars in spiral and irregular galaxies.

The brightest red supergiants were also used as standard candles to determine
 the distances to galaxies \citep{Hum1983, Kar1994}. It was assumed that red
 supergiants may be more convenient for this purpose, since they can be easily
 distinguished from compact young clusters by their color index.

When looking for bright red stars in galaxies, red dwarfs from our Galaxy may
 get on the list. Their identification was difficult to work with. One way to
 determine the type of stars was to plot their distribution over the body of
 the galaxy. The stars of the galaxy showed concentration towards the body of
 the galaxy, whereas the background stars were evenly distributed. The second
 method consisted of comparing the positions of bright red stars, which are
 likely to be massive stars in the galaxy, relative to the regions of star
 formation. As it has been already mentioned, it is more likely that the birth
 of a bright star will occur in a star-formation region, so a priority
 was given to stars near such regions. Probably, it would be
 possible to resolve the issue of separating distant and nearby stars if we
 used the results of measurements of proper motions of stars obtained by the
 Gaia \citep {Brown2021} telescopes. But so far, despite the selection
 methods, the presence of red dwarfs from our Galaxy is possible in our list
 of bright red stars.

Compact star clusters with the same color index as the brightest red stars can
only belong to old clusters. Their photometric profile differs from the profile
 of single stars, so the identification of such clusters was not particularly
 difficult. In some galaxies, bright red stars were absent. Therefore, in our
 sample the number of galaxies with blue and red stars is different.

\section{LIST OF GALAXIES} 

To search for the correlation between the luminosities of galaxies and their
 brightest stars, we used archival images of the Hubble Space Telescope (HST),
 obtained over many years of its work on various programs, mainly on
 applications for determining distances to galaxies. The galaxies in our
 sample are distributed in the $M_B$ luminosity range from $-19^{\rm m}$
 to $-8^{\rm m}$. We did not include very close galaxies (M\,31, M\,33, and
 others) in our sample, since their angular sizes were much larger than the
 field sizes of the ACS camera of HST, and it was not the purpose of this
work to compose a mosaic of many images. For some galaxies we used two fields
 to cover all the bright stars in the galaxies, but the results in the vast
 majority of cases are obtained with one field photometry. The search for
 the brightest stars is also being carried out for brighter galaxies
 ($M_B$ to $-21^{\rm m}$), which, as a rule, are located at distances
 from 10 to 20~Mpc, but the results are under interpretation.

\begin{figure*}
\centerline{\includegraphics[angle=0, scale=0.8,clip]{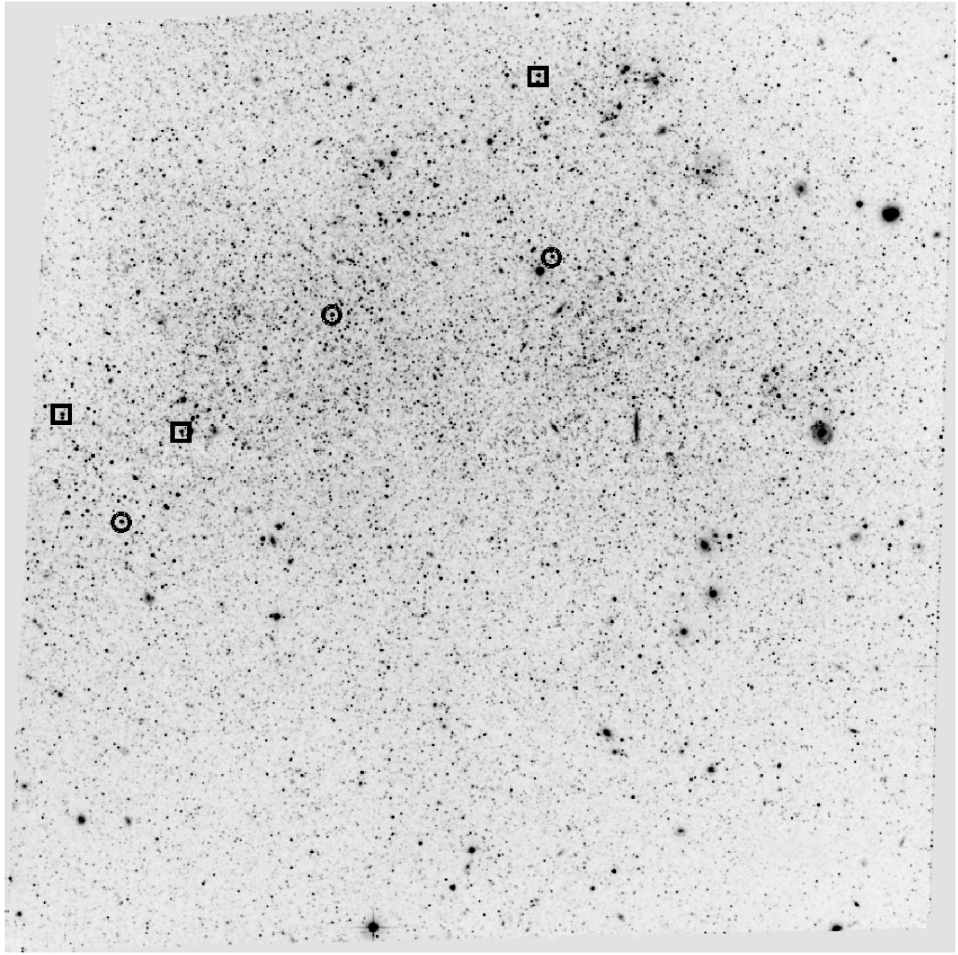}}
	\caption
{Image of the Holmberg\,I galaxy obtained with the HST telescope in the
 $F814W$ filter. The circles mark the three brightest blue stars, and the
squares mark the three red ones. The image size is $ 3\farcm5\times3\farcm5$.}
		\label{figure1}
\end{figure*}
\begin{figure}[bpt!!!]
\centerline{\includegraphics[angle=0, width=7.8cm,clip]{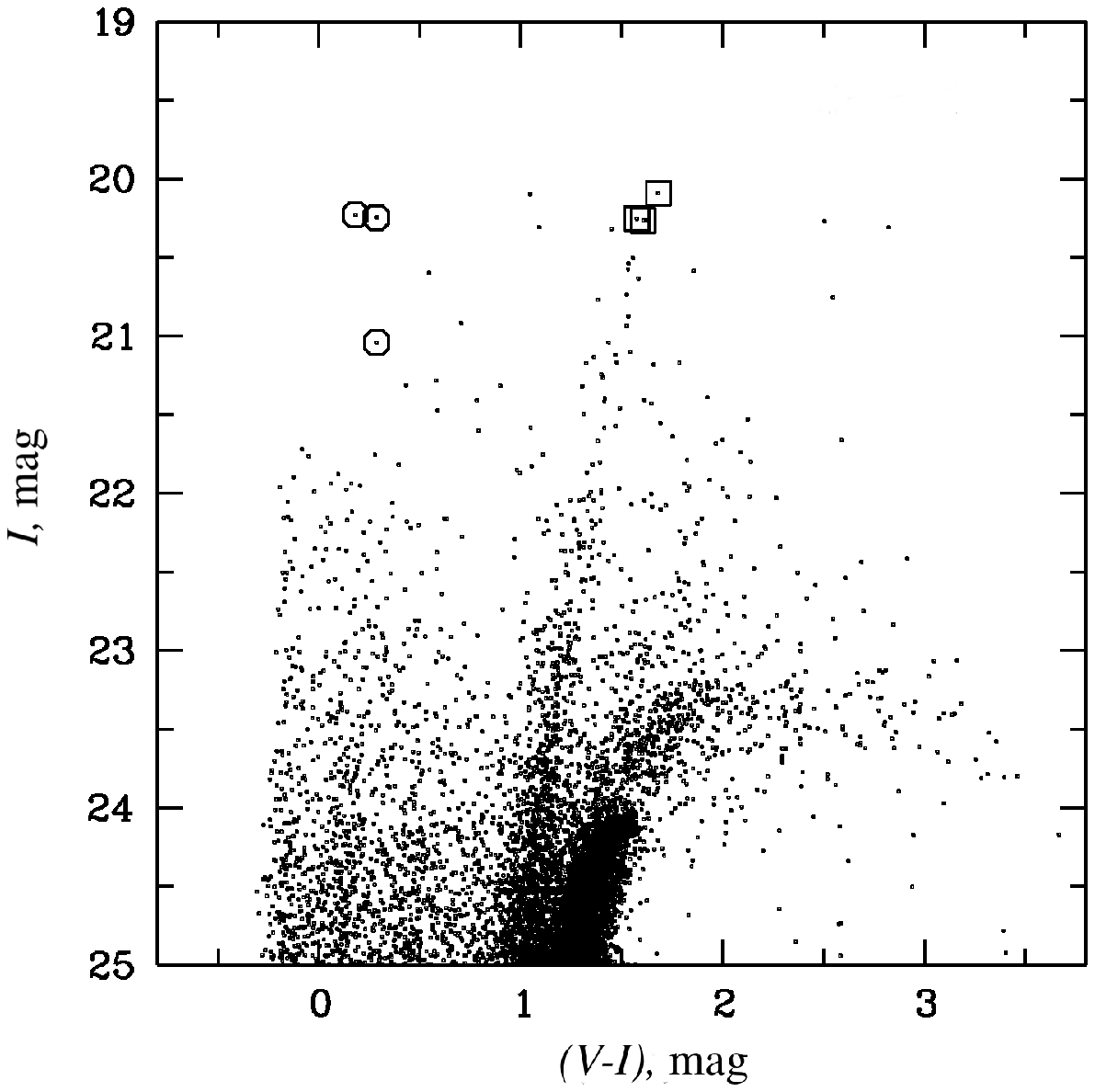}}
	\caption{
CM diagram of the Holmberg\,I galaxy obtained after photometry of the HST
telescope images. The positions of the blue (circles) and red (squares) stars
are marked.}
	\label{figure2}
\end{figure}

When determining the distances to galaxies, we tried to use no images of the
 central regions, where bright stars are located and the background is
 significantly increased, but images of the periphery of galaxies, where the
 stellar population consists of red giants and AGB stars. TRGB method
 \citet{Lee1993} was used to measure the distances, and TRGB jump position
 was determined by the Sobel function \citep{Mad1995}. Extinction value
 towards galaxies was determined based on the work \citet{Sch2011}.

Dwarf galaxies occupy only a part of the field of HST images, therefore,
 we were determining the distances from the same images that we used for
 the search for bright stars. For three galaxies, there were images only
 in the $F435W$ and $F814W$ filters, and we did the following. Based on the
 images of galaxies of a similar type, for which the images were taken in
 the $F435W$, $F606W$ and $F814W$ filters, we determined the relationship
 between the color indices $(B-I)$ and $(V-I)$. Then we derived desired values
 for the galaxies in the $V$ filter. We did not include to the sample those
 galaxies in which only a part of the galaxy with bright stars fell on the HST
 images. Since we did not select galaxies by type, the general list contains a
 variety of spiral and irregular galaxies. Results for all the galaxies in the
sample are available at {\url {https://www.sao.ru/hq/dolly/bs/table_1_150.pdf} and \url{https://www.sao.ru/hq/dolly/bs/table_2_150.pdf}}.
For example, Table~\ref{Table_itog1} shows some of them, and
Table~\ref{Table_itog2}~---contains the results of photometry of the brightest
stars obtained for the indicated galaxies. It can be added that the list is not
final and new galaxies, which are currently being worked on, will be added to it.

\begin{table*}[]
\caption{ Parameters of the investigated galaxies* }
\label{Table_itog1}
\medskip
\normalsize
\begin{tabular}{c|l|c|c|c|c|c|c|c|c}  \hline 
Number&~~~~Galaxy     &RA\,(J\,2000)   & Dec\,(J\,2000) \rule{0pt}{2.5ex} &$(m-M)$ &$D$,  Mpc  & $A_B$& $E(V-I)$& $B_t$ & $M_B$ \\ 

\hline 
(1)&  ~~~~~~(2)         &   (3)       & (4)         \rule{0pt}{2.5ex} & (5)   & (6)  & (7)     & (8)   &   (9)    & (10)\\ [2pt]
\hline
1  &AGC\,102728     &  00 00 21.4 &  +31 01 18.7 & 29.73 &  8.84&0.17 &    0.057 &  19.45& $-$10.45\\  [2pt]
2  &ESO\,349-031    &  00 08 13.5 &$-$34 34 43.6 & 27.21 &  2.77&0.04 &    0.015 &  15.69& $-$11.56\\  [2pt]
3  &NGC\,24         &  00 09 56.4 &$-$24 57 49.8 & 29.35 &  7.42&0.07 &    0.024 &  12.08& $-$17.34\\  [2pt]
4  &UGC\,288        &  00 29 04.1 &  +43 25 50.7 & 29.17 &  6.82&0.28 &    0.138 &  16.00& $-$13.45\\  [2pt]
5  &IC\,1574        &  00 43 03.7 &$-$22 14 51.4 & 27.85 &  3.73&0.06 &    0.019 &  14.47& $-$13.44\\  [2pt]
6  &DDO\,6          &  00 49 49.7 &$-$21 00 47.1 & 27.52 &  3.19&0.06 &    0.021 &  14.99& $-$12.59\\  [2pt]
7  &IC\,1613        &  01 04 47.8 &  +02 07 03.6 & 24.48 &  0.79&0.09 &    0.030 &  10.01& $-$14.56\\  [2pt]
8  &UGC\,685        &  01 07 22.4 &  +16 41 02.0 & 28.27 &  4.51&0.21 &    0.071 &  14.37& $-$14.11\\  [2pt]
9  &KKH\,6          &  01 34 51.6 &  +52 05 30.0 & 27.91 &  3.81&1.27 &    0.434 &  17.00& $-$12.18\\  [2pt]
10 &PGC\,6430       &  01 45 03.9 &$-$43 35 54.9 & 28.30 &  4.57&0.06 &    0.020 &  12.80& $-$15.55\\  [2pt]
\hline 
\multicolumn{10}{l}{\footnotesize *Coordinates of galaxies (RA and Dec),
 the magnitude of light absorption in the direction of the galaxies }\\[-0pt]
\multicolumn{10}{l}{\footnotesize ($A_B$ and $ E(V-I) $) and apparent
 magnitudes of galaxies ($B_t$) are taken from the NED and HyperLEDA  }\\[-0pt]
\multicolumn{10}{l}{\footnotesize databases, and the moduli of the distances
 $(m-M)$, the distances to galaxies $(D)$ and the absolute }\\[-0pt]
\multicolumn{10}{l}{\footnotesize luminosities of the galaxies  ($M_B$) were
 obtained by us.} \\
	\end{tabular}  
\end{table*}   


After compiling a list of galaxies, we carried out stellar photometry
 of all the images we used and built Hertzsprung-Russell diagrams (CM
 diagrams), on which the brightest blue and red stars were highlighted.
 After a visual inspection of the stars, we left three blue and three red
 brightest stars in each galaxy for statistical calculations. In
 Fig.~\ref{figure1}, for example, the Holmberg\,I galaxy is shown with
 the markings of the found brightest red and blue stars, and in
 Fig.~\ref{figure2}~--- the CM diagram of this galaxy with the selected
 stars indicated.

Based on the red giants being visible in the images, we determined the
 distances for those galaxies that are absent in the list of galaxies
 from~\citet{Tik2018}. The distances were determined by the
 TRGB method~\citep{Lee1993} in a standard way. For some galaxies,
 the distances have been obtained by the TRGB method for the first time,
 and for those galaxies for which there were already measurements, the
 obtained values coincide, within measurement error, with the values
 of other authors.

\section{STELLAR PHOTOMETRY}
The stellar photometry of HST images with the ACS camera was performed using
 the DAOPHOT \,II~\citep {Ste1987, Ste1994} software package, and the
 DOLPHOT~2.0\footnote{\url{http://americano.dolphinsim.com/dolphot/dolphot.pdf}}
 software package was used for photometry of galaxy images made with the WFC3
 camera.
Photometry of stars with both packages was carried out in a standard way. For
 DAOPHOT\,II, the procedure was described by us earlier~\citep{Tik2019}, and the
 DOLPHOT\,2.0 package was used in accordance with the recommendations of its
 author~\citet {Dol2016}. The principles of photometry with the DOLPHOT and
 DAOPHOT are the same, but there are some differences in their use.
 For example, in DAOPHOT we took single stars from the fields being studied as
 PSF stars, and in DOLPHOT we used the PSF profile library. When measuring the
 positions of TRGB jumps, both methods give similar results and no significant
 differences were found between them.

\begin{figure*}[]
	\centerline{\includegraphics[angle=0, width=14cm,clip]{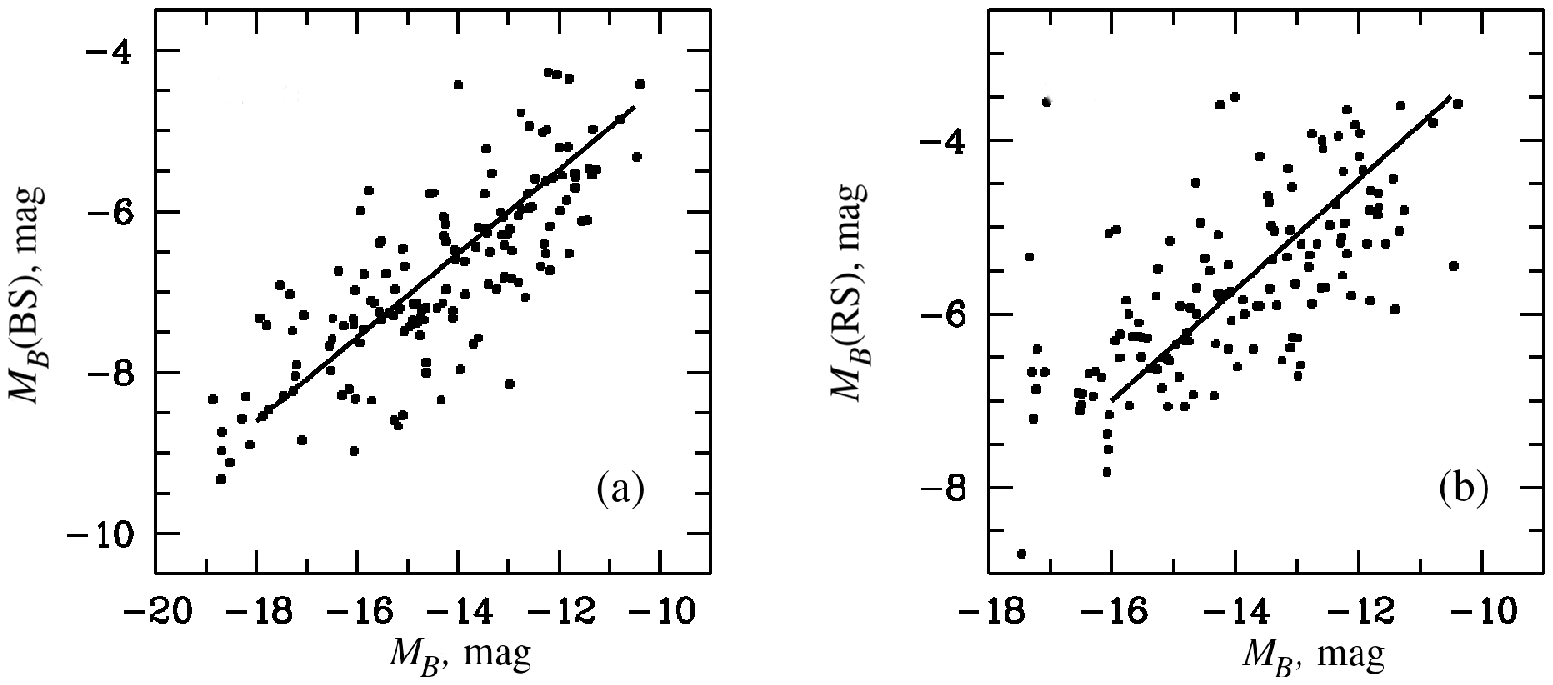}}
	\caption{Diagrams of the relationship between the luminosity of galaxies
 and the luminosity of their brightest blue~(a) and red~(b) stars.}
	\label{figure3}
\end{figure*}

The photometry procedure consisted of preliminary masking of bad pixels,
 removal of cosmic particle trails, and further PSF photometry of the
 stars found. To remove non-stellar objects: star clusters, distant or compact
 galaxies, all stars were selected according to the CHI and SHARP parameters,
 which determine the shape of the photometric profile of each star being
 measured \citep{Ste1987}. The profiles of non-stellar objects differed from
 the profiles of isolated stars that we selected as PSF-stars, which made it
 possible to carry out such a selection with the lists of stars obtained by
 DAOPHOT\,II and DOLPHOT\,2.0.

HST images for several galaxies were taken with the WFPC2 camera. We followed
 the recommendations of \citet{Hol1995a, Hol1995b} for their photometry with
 DAOPHOT. Due to the lower photometric limit of the WFPC2 camera compared to
 the ACS and WFC3 cameras, the galaxies for the WFPC2 images were taken
 are at closer distances than the rest of the galaxies.

Photometry results in the form of average stellar magnitudes and color indices
 are presented at the site~{\url{https://www.sao.ru/hq/dolly/bs/table_2_150.pdf}}
 and for some of the sample galaxies are given in Table~\ref{Table_itog2} for example.

\begin{table*}[]
\caption{Photometry results of the brightest stars in galaxies } 
\label{Table_itog2}
\medskip
\normalsize 
\begin{tabular}{c|l|c|r|c|r|c|c|c|c} 
	\hline
Number&\multicolumn{1}{c|}{Galaxy} &$m_V^{\rm 1B}$ &$(V-I)^{\rm 1B}$ &$\langle m_V^{\rm 3B}\rangle$ \rule{0pt}{2.5ex}
 &$\langle (V-I)^{\rm 3B}\rangle$ &$m_V^{\rm 1R}$  & $(V-I)^{\rm 1R}$ &$\langle m_V^{\rm 3R}\rangle$
 &$\langle (V-I)^{\rm 3R}\rangle$ \\
\hline 
(1)&    \multicolumn{1}{c|}{(2)} & (3)         & (4)~~~~~     & (5)   & (6)~~~~~~  & \rule{0pt}{2.5ex}(7) & (8)      &(9)    &(10)  \\ \hline 
1  &AGC\,102728     &      24.398 &    0.101    &   24.535    &       0.099      &  23.333  &1.513 \rule{0pt}{2.5ex}      &     24.404  &   1.583      \\ [2pt]
2  &ESO\,349-031    &      20.711 & $-$0.049    &   21.125    &    $-$0.022      &  21.184  &1.529       &     22.055  &   1.600      \\ [2pt]
3  &NGC\,24         &      22.214 &    0.085    &   22.380    &       0.119      &  23.671  &1.451       &     24.057  &   1.493      \\ [2pt]
4  &UGC\,288        &      22.914 & $-$0.088    &   23.209    &       0.031      &  24.410  &1.668       &     24.709  &   1.611      \\ [2pt]
5  &IC\,1574        &      22.550 &    0.290    &   22.671    &       0.113      &  22.084  &1.500       &     22.182  &   1.503      \\ [2pt]
6  &DDO\,6          &      22.441 &    0.267    &   22.634    &       0.244      &  23.235  &2.855       &     23.566  &   2.881      \\ [2pt]
7  &IC\,1613        &      18.787 &    0.121    &   18.774    &    $-$0.060      &  19.297  &1.634       &     19.595  &   1.575      \\ [2pt]
8  &UGC\,685        &      20.401 &    0.016    &   21.193    &       0.045      &  22.742  &1.701       &     22.992  &   1.560      \\ [2pt]
9  &KKH\,6          &      22.124 &    0.711    &   22.689    &       0.718      &  22.845  &2.566       &     23.562  &   2.278      \\ [2pt]
10 &PGC\,6430       &      21.774 &    0.260    &   21.955    &       0.150      &  21.888  &1.500       &     22.087  &   1.507      \\ [2pt]
\hline                                                                                                                  
                                       
\multicolumn{10}{l}{\footnotesize 	In the columns of the Table: }  \\[-0pt]
\multicolumn{10}{l}{\footnotesize 	$m_V^{\rm 1B}$~---  apparent magnitude of the brightest blue star,  
	                                 $(V-I)^{\rm 1B}$~--- its color index,}\\[-0pt]
\multicolumn{10}{l}{\footnotesize       $\langle m_V^{\rm 3B}\rangle$~--- the average apparent magnitude
 of the three brightest blue stars,  }\\[-0pt]
\multicolumn{10}{l}{\footnotesize       $\langle (V-I)^{\rm 3B}\rangle$~--- the average color index of
 these stars, }\\[-0pt]
\multicolumn{10}{l}{\footnotesize 	$m_V^{\rm 1R}$~--- apparent magnitude of the brightest red star,
	                                                  $(V-I)^{\rm 1R}$~--- its color index}\\[-0pt]
\multicolumn{10}{l}{\footnotesize       $\langle m_V^{\rm 3R}\rangle$ ~--- the average apparent magnitude
 of the three brightest red stars,}\\[-0pt]
\multicolumn{10}{l}{\footnotesize       $\langle (V-I)^{\rm 3R}\rangle$~---  the average color index of
 these stars. }\\[-0pt]
	\end{tabular} 
\end{table*}                                                                                                        


\section{GALAXIES AND THEIR BRIGHT STARS} 

Based on the data on the parameters of the 150 studied galaxies in the
 sample and the results of photometry of their brightest stars (see Tables~
 \ref{Table_itog1} and \ref{Table_itog2}), we have constructed diagrams of
 dependences between the luminosities of $M_B$ galaxies and the average absolute 
 luminosities of three brightest blue and red stars in the $V$ band,
 $M_V(\rm{BS})$ and $M_V(\rm{RS})$ respectively (Fig.~\ref {figure3}). 
 The indicated quantities are determined from the relation:
 $$ M_V({\rm BS})= \langle m_V^{\rm 3B}\rangle - A_V -(m-M),\eqno(1)$$
 $$ M_V({\rm RS})= \langle m_V^{\rm 3R}\rangle - A_V-(m-M).\eqno(2)$$
   It can be seen
 that for low and medium luminosity galaxies ($-19^{\rm m}<M_B<-10^{\rm m}$)
 for blue stars and ($-17^{\rm m} <M_B<-10^{\rm m}$) for red stars, there are
 linear dependences between the indicated values, (3)~--- for blue and
 (4)~--- for red:
 $$M_V(\rm{BS}) = 0.471M_B + 0.059, \sigma=0.38,\eqno(3)$$
 $$M_V(\rm{RS}) = 0.539M_B + 1.854, \sigma=0.42.\eqno(4)$$

As it was pointed above, the brightest and most massive stars should be
 observed in galaxies with very low metallicity \citep{Yus2013}. Since
 low-metal galaxies can only be dwarf galaxies, in which the star formation
 process consists of individual irregular flares, such galaxies should have
 a maximum spread in the luminosities of stars, from the brightest hypergiants
 to normal main sequence stars. However, in the diagram in Fig.~\ref{figure3}a 
 we do not see that there are very bright blue stars in faint galaxies
 \mbox{($-13{\rm^m}<M_B<-10{\rm^m}$).}
On the same diagram for the region of bright galaxies  \mbox{($-18^{\rm m}<M_B<-13^{\rm m}$),}
 we see that many galaxies have bright stars, and for some galaxies,
 stars have a brightness significantly higher than the average value from
 equation~3 (straight line in Fig.~\ref{figure3}a).

\begin{figure*}[]
\centerline{\includegraphics[angle=0, width=13cm,clip]{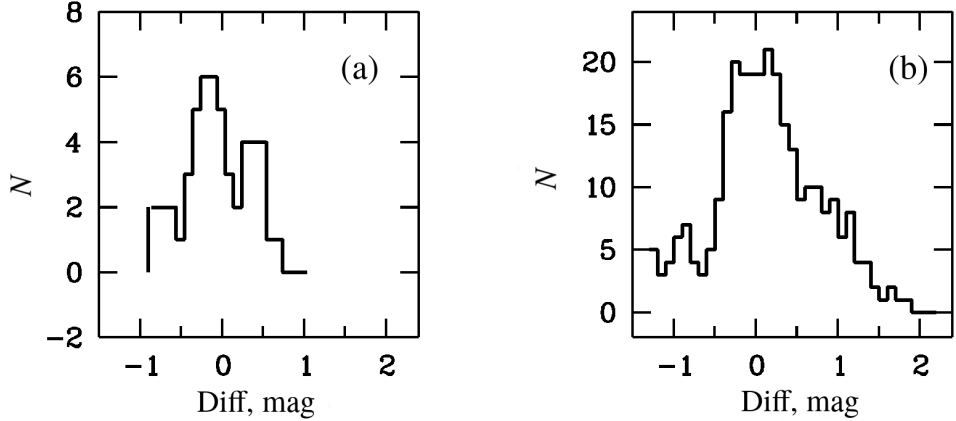}}
\caption{Distribution of the number of galaxies relative to the linear
 dependence for blue stars Fig.~\ref{figure3}. The smallest galaxies
 ($M_B>-13^{\rm m}$) deviate from the mean only within the statistical
 values~(a). For galaxies of medium luminosity ($-18^{\rm m} <M_B <-13^{\rm m}$),
 asymmetry~(b) is visible in the distribution of their abundance, that is,
 the number of galaxies whose blue stars have a brightness increased to two
 magnitudes relative to the average brightness of such stars.}
\label{figure4}
\end{figure*}

More clearly the observed regularity in the increase in the luminosity of
 the brightest stars in brighter galaxies is shown in Fig.~\ref{figure4},
 which shows the distributions of the number of galaxies with different
 luminosities relative to the above linear dependence~(3) for blue stars.
 Low luminosity galaxies \mbox{($-13^{\rm m}<M_B <-10^{\rm m}$)} have a simple
 statistical scatter relative to dependence~(3), that is, no excess is visible
 in the number of galaxies with stars of increased brightness. But in
 galaxies of average luminosity \mbox{($-18^{\rm m}<M_B<-13^{\rm m}$)} there
 is a good noticeable asymmetry in distribution. An increase in the number
 of galaxies is seen, where the stars are brighter relative to the
 dependence~(3). The excess of galaxies with brighter stars is what causes
 the observed distribution asymmetry.

This result has a simple explanation. The faintest galaxies are galaxies
 without intense star formation and one should not expect the
 appearance of very bright stars in such galaxies. These galaxies have
 insignificant masses of hydrogen and low-mass clusters are formed in the
 process of star formation. Therefore, the birth of a supermassive brightest
 star in such a galaxy is unlikely.

More bright and massive galaxies \mbox{($-18^{\rm m}<M_B<-13^{\rm m}$)} have enough
 hydrogen, and their metallicity is still low compared to the metallicity
 of massive spiral galaxies, so you can expect the appearance of supermassive
 stars in such galaxies. With active star formation processes, the luminosity
 of galaxies increases, especially in the blue region, due to the appearance
 of young blue stars, therefore, during an outbreak of star formation, when
 the birth of bright massive stars can be expected, the galaxy itself has an
 increased brightness. This can be seen in the diagram in Fig.~\ref{figure4},
 where the distribution asymmetry, that is, the appearance of high-brightness
 stars is observed in galaxies with $-18^{\rm m} <M_B<-13^{\rm m} $.

\section{REMARK ABOUT THE METHOD OF DETERMINING THE DISTANCES BY THE BRIGHT STARS }

\begin{figure}[]
	\centerline{\includegraphics[angle=0, width=6.9cm,clip]{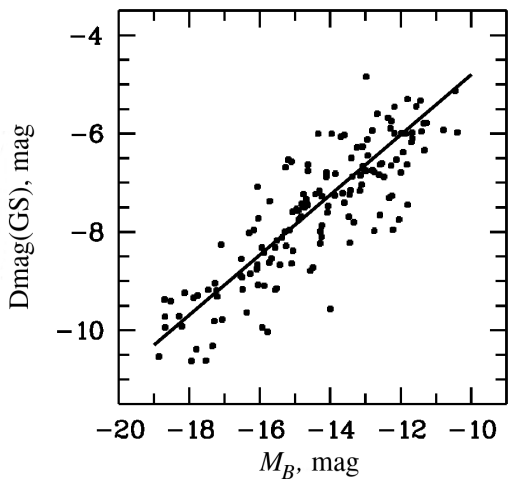}}
	\caption{Diagram of the relationship between the absolute magnitudes of galaxies and the difference
 between the apparent magnitudes of galaxies and the average brightness of three blue stars. The diagram
 allows determining the distance to galaxies based on the photometry of their blue stars. }
	\label{figure5}
\end{figure}

 The diagrams in Fig.~\ref{figure3}
 clearly show the dependences between the luminosities of galaxies and stars,
 but in this form they can not be used to determine the distance. The diagram
 in Fig.~\ref{figure5} shows the relationship between the absolute luminosity
 of $M_B$ galaxies and the difference between the apparent stellar magnitude
 of the $B_t$ galaxy and the average apparent stellar magnitude of three blue
 stars: $${\rm Dmag (GS)} = B_t - \langle m_V^{\rm 3B} \rangle.$$
The presented diagram makes it possible to calculate the absolute luminosity
 of the galaxy $M_B$ from the difference in the magnitudes of the galaxy and
 the stars and then determine the distance to it. The following dependencies
 were obtained:
$$B_t -\langle m_V^{\rm 3B}\rangle = 0.611\times M_B + 1.311,\sigma=0.41,\eqno(5)$$
$$M_B = \dfrac{(B_t - \langle m_V^{\rm 3B}\rangle - 1.311)}{ 0.611}.\eqno(6)$$

Thus, based on the photometry of blue stars and the total stellar
 magnitude of the galaxy, we can use equation (6) to determine the absolute
 magnitude of the galaxy ($M_B$) and find the value of the distance modulus:
 $$(m-M)=B_t-M_B.$$ Then the distance is determined from the equation:
$$ \log D = (m-M)/5 + 1~({\rm pc}).$$ It should be noted that the values
 of $B_t$ and $\langle m_V^{\rm 3B} \rangle$ were corrected for extinction
 in our Galaxy based on work~\citet{Sch2011}.

It is interesting to note that when choosing galaxies at different distances,
 we did not find the influence of the distance effect, that is, for galaxies
 located closer than 7~Mpc, the deviation of points from the dependence is
 practically the same as for galaxies at distances up to 20~Mpc. This means
 that stars and clusters can be separated in HST images up to the specified
 distance.

Similarly to blue stars, red stars can be used, however, the dependence has
 a stronger deviation of points and the accuracy of distance calculations
 will be worse. The main advantage of using blue stars is that blue stars
 are easier to distinguish from other stars, and in addition,
 in small dwarf galaxies with star-formation regions there are more blue stars
 than red stars, so choosing blue stars increases the measurement accuracy.

The accuracy of the method can be improved if additional parameters are
 introduced into the dependence (Fig.~\ref {figure3}). For example, to use
 the color indices of galaxies or flux values in the H$\alpha$ filter, which
 reflect the intensity of star formation processes. Moreover, one can use the
 infrared luminosity of galaxies, which more accurately determines the value
 of their mass.
We have not used such opportunities, as it is beyond the scope of our work,
 and present the results of our measurements without any correction.

\section{CONCLUSIONS }
 
The brightest blue and red stars have been identified in 150 irregular and
 low-mass spiral galaxies based on photometry of the Hubble Space Telescope
 images. Distances were determined for all the galaxies using the TRGB method.

 When comparing the average absolute luminosities of the three brightest blue stars and
 the absolute luminosities of their host galaxies, a linear relationship was found:
 $$M_B = (B_t - \langle m_V^{\rm 3B}\rangle - 1.311) / 0.611,$$
 which can be used to determine distances to galaxies. This method  can find
 its application for irregular and low-massive distant spiral galaxies,
 in which the red giants are not visible.

The luminosity diagram of galaxies and blue stars (Fig.~\ref{figure3}a)
 shows that there are no bright stars in faint galaxies. The sample of
 galaxies ($-12\fm5<M_B<-10{\rm^m}$) includes 31 galaxies with an average
 luminosity of \mbox{$M_B=-11\fm9 $.} The total luminosity of these galaxies
 is approximately $M_B = -15\fm5$. All faint galaxies on the diagram do not
 have a single bright blue star ($M_V < -7^{\rm m}$), while galaxies with luminosity
 \mbox{$-16{\rm ^m} <M_B <-15{\rm ^m}$} have such stars. Earlier it was
 believed that the appearance of brighter stars in more massive galaxies is
 explained only by a larger number of stars than in small galaxies. Based on
 the absence of bright stars in 31 faint galaxies, the total luminosity of
 which corresponds to the luminosity of one average galaxy containing bright
 stars, it can be concluded not about the statistical relationship between
 the greater number of stars and the probability of the appearance of bright
 hypergiants, but about the action of physical processes on the mechanism
 of formation of bright massive stars in galaxies of different mass.

The absence of bright stars in dwarf galaxies $ M_B>-13{\rm ^m} $ allows us
 not to consider them as candidate objects when searching for the brightest
 and supermassive stars, although it is these galaxies that have the lowest
 metallicity, at which stars with maximum masses should be born.

It is possible that at present there are no conditions at all in any galaxies
 for the birth of supermassive stars. Dwarf galaxies have too small masses,
 and more massive galaxies have acquired an increased metallicity of gas
 during evolution, which prevents the formation of supermassive stars. That
 is, favorable conditions for the formation of such stars could exist only
 in the initial stage of the formation of galaxies.
\section*{FINANCING}
The study was financially supported by the Russian Foundation for Basic
Research and the National Science Foundation of Bulgaria as a part of the
scientific project N19--52--18007 and Grant KP--06--Russin--09/2019.\\

\section*{ACKNOWLEDGMENT}

Based on observations with the NASA/ESA Hubble Space Telescope, obtained at the
Space Telescope Science Institute, which is operated by AURA, Inc. under contract No. NAS5-26555. These 
observations are associated with the proposals 5091, 5375, 5397, 5427, 5915, 5971, 5972,
6431, 6549, 6584, 6695, 6865, 7202, 7496, 8059, 8122, 8192, 8199, 8584,
8601, 9042, 9086, 9162, 9765, 9771, 9774, 9820, 10182, 10210, 10235, 10402,
10433, 10438, 10505, 10523, 10585, 10605, 10696, 10765,  10877, 10885,
10889, 10905, 10915, 10918, 11229,  11307, 11360, 11575, 11718, 11986, 12196,
12546, 12878, 12880, 12902, 12968, 13357, 13364, 13442, 13750, 14678, 15133,
15243, 15275, 15564, 16075.




\end{document}

%% file: sao_cmd_author.tex
\def\saoname{Special Astrophysical Observatory,  Russian Academy of Sciences,
              Nizhnii Arkhyz, 369167 Russia}

\def\squareforqed{\hbox{\rlap{$\sqcap$}$\sqcup$}}

\def\sq{\ifmmode\squareforqed\else{\unskip\nobreak\hfil
\penalty50\hskip1em\null\nobreak\hfil\squareforqed
\parfillskip=0pt\finalhyphendemerits=0\endgraf}\fi}

\def\utw{\smash{\rlap{\lower5pt\hbox{$\sim$}}}}

\def\udtw{\smash{\rlap{\lower6pt\hbox{$\approx$}}}}

\def\fm{\hbox{$\,.\!\!^{\rm m}$}}

\def\farcm{\hbox{$\,.\mkern-4mu^\prime$}}

\def\diameter{{\ifmmode\mathchoice
{\ooalign{\hfil\hbox{$\displaystyle/$}\hfil\crcr
{\hbox{$\displaystyle\mathchar"20D$}}}}
{\ooalign{\hfil\hbox{$\textstyle/$}\hfil\crcr
{\hbox{$\textstyle\mathchar"20D$}}}}
{\ooalign{\hfil\hbox{$\scriptstyle/$}\hfil\crcr
{\hbox{$\scriptstyle\mathchar"20D$}}}}
{\ooalign{\hfil\hbox{$\scriptscriptstyle/$}\hfil\crcr
{\hbox{$\scriptscriptstyle\mathchar"20D$}}}}
\else{\ooalign{\hfil/\hfil\crcr\mathhexbox20D}}%
\fi}}

\newcommand{\ab}{Astrophysical Bulletin }

\newcommand{\aaa}{Astron. and Astrophys. }
\newcommand{\aap}{Astron. and Astrophys. }

\newcommand{\aj}{Astron.~J. }

\newcommand{\mnras}{Monthly Notices Royal Astron. Soc. }

\newcommand{\pasp}{Publ. Astron. Soc. Pacific }












%% file: Tikhonov.bbl
\begin{thebibliography}{32}
	\providecommand{\natexlab}[1]{#1}
	
	\bibitem[{Antipova} et~al.(2020)]{Antipova2020}
	A.~V. {Antipova}, D.~I. {Makarov}, and L.~N. {Makarova}, Astrophysical Bulletin
	\textbf{75}~(2), 93 (2020).
	
	\bibitem[{Bestenlehner} et~al.(2020)]{Bes2020}
	J.~M. {Bestenlehner}, P.~A. {Crowther}, S.~M. {Caballero-Nieves}, et~al.,
	\mnras \textbf{499}~(2), 1918 (2020).
	
	\bibitem[Brown(2021)]{Brown2021}
	A.~G. Brown, Annual Review of Astronomy and Astrophysics \textbf{59}~(1),
	59–115 (2021).
	
	\bibitem[{Crowther} et~al.(2010)]{Cro2010}
	P.~A. {Crowther}, O.~{Schnurr}, R.~{Hirschi}, et~al., \mnras \textbf{408}~(2),
	731 (2010).
	
	\bibitem[{de Vaucouleurs}(1978)]{Vau1978}
	G.~{de Vaucouleurs}, \apj \textbf{224}, 710 (1978).
	
	\bibitem[{de Wit} et~al.(2004)]{deW2004}
	W.~J. {de Wit}, L.~{Testi}, F.~{Palla}, et~al., \aap \textbf{425}, 937 (2004).
	
	\bibitem[{Dolphin}(2016)]{Dol2016}
	A.~{Dolphin}, {DOLPHOT: Stellar photometry} (2016).
	
	\bibitem[{Holmberg}(1950)]{Hol1950}
	E.~{Holmberg}, Meddelanden fran Lunds Astronomiska Observatorium Serie II
	\textbf{128}, 5 (1950).
	
	\bibitem[{Holtzman} et~al.(1995{\natexlab{a}})]{Hol1995b}
	J.~A. {Holtzman}, C.~J. {Burrows}, S.~{Casertano}, et~al., \pasp \textbf{107},
	1065 (1995{\natexlab{a}}).
	
	\bibitem[{Holtzman} et~al.(1995{\natexlab{b}})]{Hol1995a}
	J.~A. {Holtzman}, J.~J. {Hester}, S.~{Casertano}, et~al., \pasp \textbf{107},
	156 (1995{\natexlab{b}}).
	
	\bibitem[{Hubble}(1936)]{Hub1936}
	E.~{Hubble}, \apj \textbf{84}, 270 (1936).
	
	\bibitem[{Humphreys}(1983)]{Hum1983}
	R.~M. {Humphreys}, \apj \textbf{269}, 335 (1983).
	
	\bibitem[{Karachentsev} and {Tikhonov}(1994)]{Kar1994}
	I.~D. {Karachentsev} and N.~A. {Tikhonov}, \aap \textbf{286}, 718 (1994).
	
	\bibitem[{Lee} et~al.(1993)]{Lee1993}
	M.~G. {Lee}, W.~L. {Freedman}, and B.~F. {Madore}, \apj \textbf{417}, 553
	(1993).
	
	\bibitem[{Lundmark}(1919)]{Lun1919}
	K.~{Lundmark}, Astronomische Nachrichten \textbf{209}~(24), 369 (1919).
	
	\bibitem[{Madore} and {Freedman}(1995)]{Mad1995}
	B.~F. {Madore} and W.~L. {Freedman}, \aj \textbf{109}, 1645 (1995).
	
	\bibitem[{Richardson} and {Mehner}(2018)]{Ric2018}
	N.~D. {Richardson} and A.~{Mehner}, Research Notes of the American Astronomical
	Society \textbf{2}~(3), 121 (2018).
	
	\bibitem[{Riess} et~al.(2016)]{Riess1996}
	A.~G. {Riess}, L.~M. {Macri}, S.~L. {Hoffmann}, et~al., VizieR Online Data
	Catalog J/ApJ/826/56 (2016).
	
	\bibitem[{Salpeter}(1955)]{Sal1955}
	E.~E. {Salpeter}, \apj \textbf{121}, 161 (1955).
	
	\bibitem[{Sandage}(1958)]{San1958}
	A.~{Sandage}, \apj \textbf{127}, 513 (1958).
	
	\bibitem[{Sandage} and {Tammann}(1974)]{San1974}
	A.~{Sandage} and G.~A. {Tammann}, \apj \textbf{194}, 223 (1974).
	
	\bibitem[{Schlafly} and {Finkbeiner}(2011)]{Sch2011}
	E.~F. {Schlafly} and D.~P. {Finkbeiner}, \apj \textbf{737}~(2), 103 (2011).
	
	\bibitem[{Stetson}(1987)]{Ste1987}
	P.~B. {Stetson}, \pasp \textbf{99}, 191 (1987).
	
	\bibitem[{Stetson}(1994)]{Ste1994}
	P.~B. {Stetson}, \pasp \textbf{106}, 250 (1994).
	
	\bibitem[{Tehrani} et~al.(2019)]{Teh2019}
	K.~A. {Tehrani}, P.~A. {Crowther}, J.~M. {Bestenlehner}, et~al., \mnras
	\textbf{484}~(2), 2692 (2019).
	
	\bibitem[{Tikhonov} et~al.(2021)]{Tik2021}
	N.~{Tikhonov}, O.~{Galazutdinova}, O.~{Sholukhova}, et~al., Research in
	Astronomy and Astrophysics \textbf{21}~(4), 098 (2021).
	
	\bibitem[{Tikhonov}(2018)]{Tik2018}
	N.~A. {Tikhonov}, Astrophysical Bulletin \textbf{73}~(1), 22 (2018).
	
	\bibitem[{Tikhonov} et~al.(2019)]{Tik2019}
	N.~A. {Tikhonov}, O.~A. {Galazutdinova}, and G.~M. {Karataeva}, Astrophysical
	Bulletin \textbf{74}~(3), 257 (2019).
	
	\bibitem[{Tully} and {Fisher}(1977)]{TF1977}
	R.~B. {Tully} and J.~R. {Fisher}, \aap \textbf{500}, 105 (1977).
	
	\bibitem[{Willick}(1996)]{Willick1996}
	J.~A. {Willick}, arXiv e-prints astro-ph/9610200 (1996).
	
	\bibitem[{Wofford} et~al.(2020)]{Wof2020}
	A.~{Wofford}, V.~{Ram{\'\i}rez}, J.~C. {Lee}, et~al., \mnras \textbf{493}~(2),
	2410 (2020).
	
	\bibitem[{Yusof} et~al.(2013)]{Yus2013}
	N.~{Yusof}, R.~{Hirschi}, G.~{Meynet}, et~al., \mnras \textbf{433}~(2), 1114
	(2013).
	
\end{thebibliography}
